\documentclass[a4paper,conference]{IEEEtran}

\usepackage{cite}      

\usepackage{graphicx}  

\usepackage{subfigure} 
\usepackage{fancyheadings}
\begin{document}

\title{Projecting SPH Particles in Adaptive Environments}

\author{\IEEEauthorblockN{Josh Borrow and Ashley J. Kelly}
\IEEEauthorblockA{Institute for Computational Cosmology\\
Durham University, Department of Physics\\
Durham, United Kingdom\\
joshua.borrow@durham.ac.uk}}


\maketitle

\begin{abstract}
The reconstruction of a smooth field onto a fixed grid is a necessary step for direct comparisons to various real-world observations. Projecting SPH data onto a fixed grid becomes challenging in adaptive environments, where some particles may have smoothing lengths far below the grid size, whilst others are resolved by thousands of pixels. In this paper we show how the common approach of treating particles below the grid size as Monte Carlo tracers of the field leads to significant reconstruction errors, and despite good convergence properties is unacceptable for use in synthetic observations in astrophysics. We propose a new method, where particles smaller than the grid size are `blitted' onto the grid using a high-resolution pre-calculated kernel, and those close to the grid size are subsampled, that allows for converged predictions for projected quantities at all grid sizes.
\end{abstract}

\section{Introduction}
\label{sec:introduction}

SPH is well known for its adaptivity; by tracing mass rather than volume the method can accurately capture the large dynamic range (of a factor $10^{11}$ in density, and $10^4$ in smoothing length, in cosmological galaxy formation simulations, see \cite{Borrow2018}) present in astrophysical phenomena, and is hence a natural choice for such simulations. The post-processing of SPH data in adaptive environments, however, presents a challenge due to the unpredictable structure of the underlying particle fields.

One post-processing case of interest, particularly within the astrophysics community, is the projection of three dimensional data onto a fixed two-dimensional grid. In astrophysics, projected quantities naturally correspond to those that are observable; for instance, it is significantly easier to measure the gas column density of a galaxy than it is to probe the three dimensional density structure. With projected quantities being instrumental in the connection of simulated datasets and the real Universe, obtaining accurate projected reconstructions of density (and other, e.g. temperature) fields from SPH data is vital. In specialised post-processing fields such as radiative transfer convergence and accuracy are a subject of regular discussion (e.g. \cite{Yang2020}), but for projected SPH quantities it is generally assumed that basic algorithms will suffice.

Typical solutions to the `projection problem' do not guarantee converged results in highly adaptive situations, particularly in the case where the smoothing length is of a similar order to the cell size that is chosen for projection. In this context, convergence refers to producing a result at a given grid resolution that corresponds to a resolution where all particles are resolved by a large number of cells, down-sampled to the lower resolution. Past work (e.g. \cite{Koepferl2017, Petkova2018}) introduced the concept of using a Voronoi decomposition to ensure that each pixel is assigned the correct fraction of each particle. This procedure, however, is computationally and conceptually complex.

We consider various methods to project SPH data to a fixed grid, from basic kernel interpolation, to `blitting' kernels to the fixed grid, and subsampling the kernel. There are many software packages available to perform this deposition, including (but not limited to) those described in \cite{Price2007, Turk2011, Benitez-Llambay2015}. The open-source package used in this work is {\tt swiftsimio} \cite{Borrow2020a}, used for reading and visualising data produced by the {\tt SWIFT} \cite{Schaller2016} code, and is the first package that we are aware of that makes subsampled techniques publicly available.

\section{Basic SPH Projection}

As noted in the introduction, there are many ways to project SPH data. The most common approach uses a pre-projected version of the kernel and for each particle loops over all pixels, calculating the kernel contribution from each particle to each pixel. It is important to ensure that the projected kernel corresponds directly to the same kernel used in the simulation, projected along one axis, to get the most accurate reconstruction. Here, for simplicity, we ignore the limb-brightening effects associated with this and simply use the two-dimensional version of the Wendland-C2 kernel,
\begin{equation}
  W(r, h) = C \cdot \max\left[\left(1 - \frac{r}{h}\right)^4 \left(1 + 4 \frac{r}{h}\right), 0\right],
  \label{eqn:kernel}
\end{equation}
with $r$ the inter-particle separation (or here the separation between the pixel sampling point and the particle), and $h$ the smoothing length of the particle. We take the ratio between the smoothing length and kernel extent $H/h = 1.897367$, and $C=7/\pi$ in accordance with \cite{Wendland1995, Dehnen2012}\footnote{The specific implementation of the kernel used here is defined in {\tt swiftsimio.visualisation.projection\_backends.kernels}.}.

\begin{figure}
  \centering
  \includegraphics[width=\columnwidth]{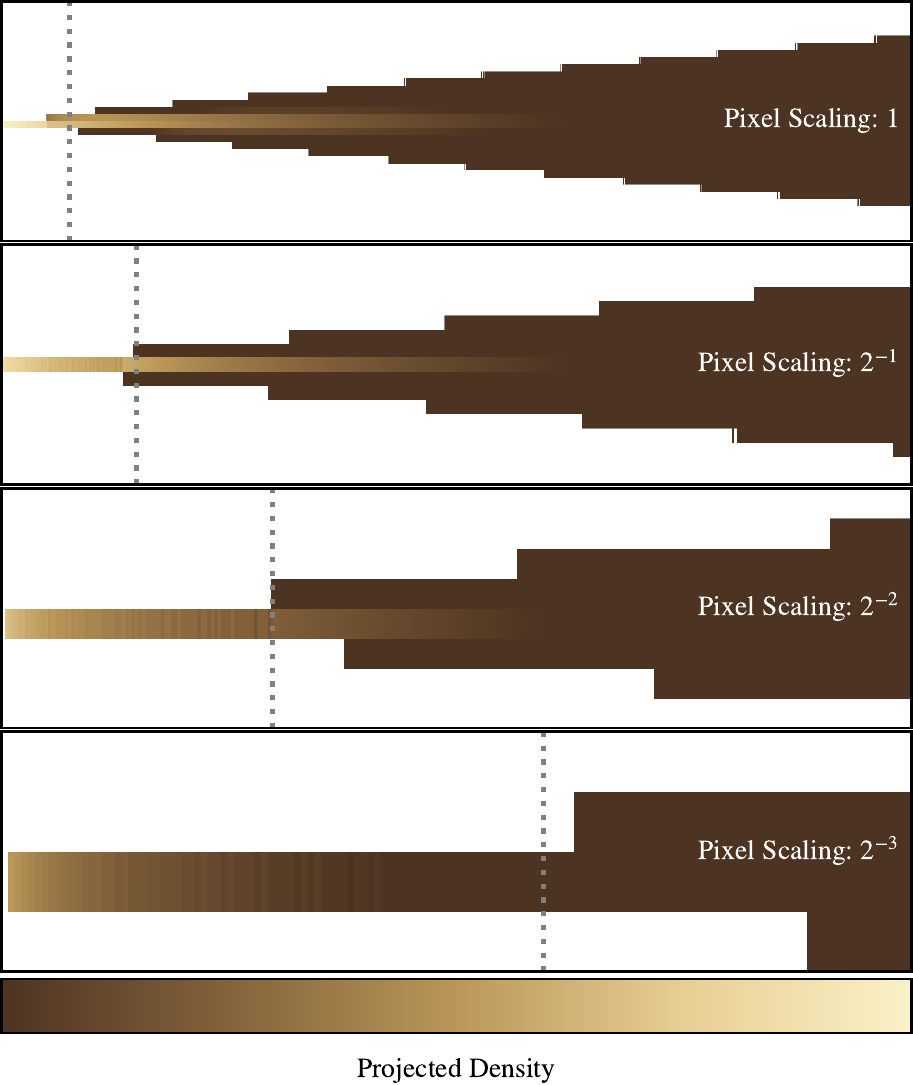}
  \caption{Projected densities of a line of particles with a linearly increasing smoothing length. White pixels correspond to areas with exactly zero density (i.e. these pixels are not reached by the kernel extent of the particles). Each panel shows a different grid resolution, with the top containing 1095 pixels in the horizontal direction, the second row 547, and so on. The left hand side, where the pixels are much larger than the kernel extent of the particles, shows striped patterns as the uncorrected visualisation method struggles to capture the smooth gradient with such poor sampling. The pixels have been stretched vertically to more clearly display the gradients. The grey dotted line shows where the kernel extent of the particles is approximately equal to half a pixel width.}
  \label{fig:gradient}
\end{figure}

Algorithmically, such a scheme (described below as `Original') looks like:
\begin{enumerate}
  \item Start with a pixel grid full of zeroes, with the pixels of size ${\rm d}x$ along one axis.
  \item If the particle has a kernel extent $H < 0.5 {\rm d}x$, add all of its contribution onto the nearest cell centre, as this particle is `unresolved', otherwise:
  \begin{enumerate}
    \item Take all particle positions $\mathbf{r}_{\rm part}$ and particle  smoothing lengths $h$, and find which pixels this cell overlaps.
    \item Loop over the cell centers, $\mathbf{r}_{\rm pix}$ and evaluate the projected kernel $W(|\mathbf{r}_{\rm part} - \mathbf{r}_{\rm pix}|, h)$.
    \item Add this contribution onto the value of each pixel.
  \end{enumerate}
\end{enumerate}

In Fig. \ref{fig:gradient} we demonstrate this algorithm at various grid resolutions on a line of (3D) particles, projected into 2D. These are generated, from left to right, starting at an inter-particle separation of $10^{-4}$ in units of the box length, linearly increasing to $10^{-2.5}$. The smoothing length $h$ of the particles is set to be the same as the inter-particle separation.

In Fig. \ref{fig:gradient} the grey dotted line denotes the region where $H \approx 0.5 {\rm d}x$. The banding pattern to the left of this line is not a compression artefact, rather
it is the result of treating particles at this grid resolution as Monte Carlo
tracers of the field. Here, the particle spacing is approximately the same as the grid size, and so we expect the shot noise to be significant (with one or two sampling points in each pixel, the shot noise is of order 100\%). As the density increases, particularly for the larger pixels in the bottom panel, the signal to noise ratio increases and the smooth gradient is again recovered.

There is no excuse for the poor reconstruction of this smooth gradient; the problem demonstrated here is purely static. We have simply chosen to throw away everything we know about the smoothness of the field at a specific grid size and as a result have been penalised with significant reconstruction errors. This method, as we will demonstrate later, does converge to the correct answer when all particles are well resolved by the grid, but for some applications grids of that size are prohibitive and large pixel sizes (relative to the particle spacing) may be required for direct comparison to real-world observations.
\section{Constructing Converged Projections}

\begin{figure*}
  \centering
  \includegraphics[width=\textwidth]{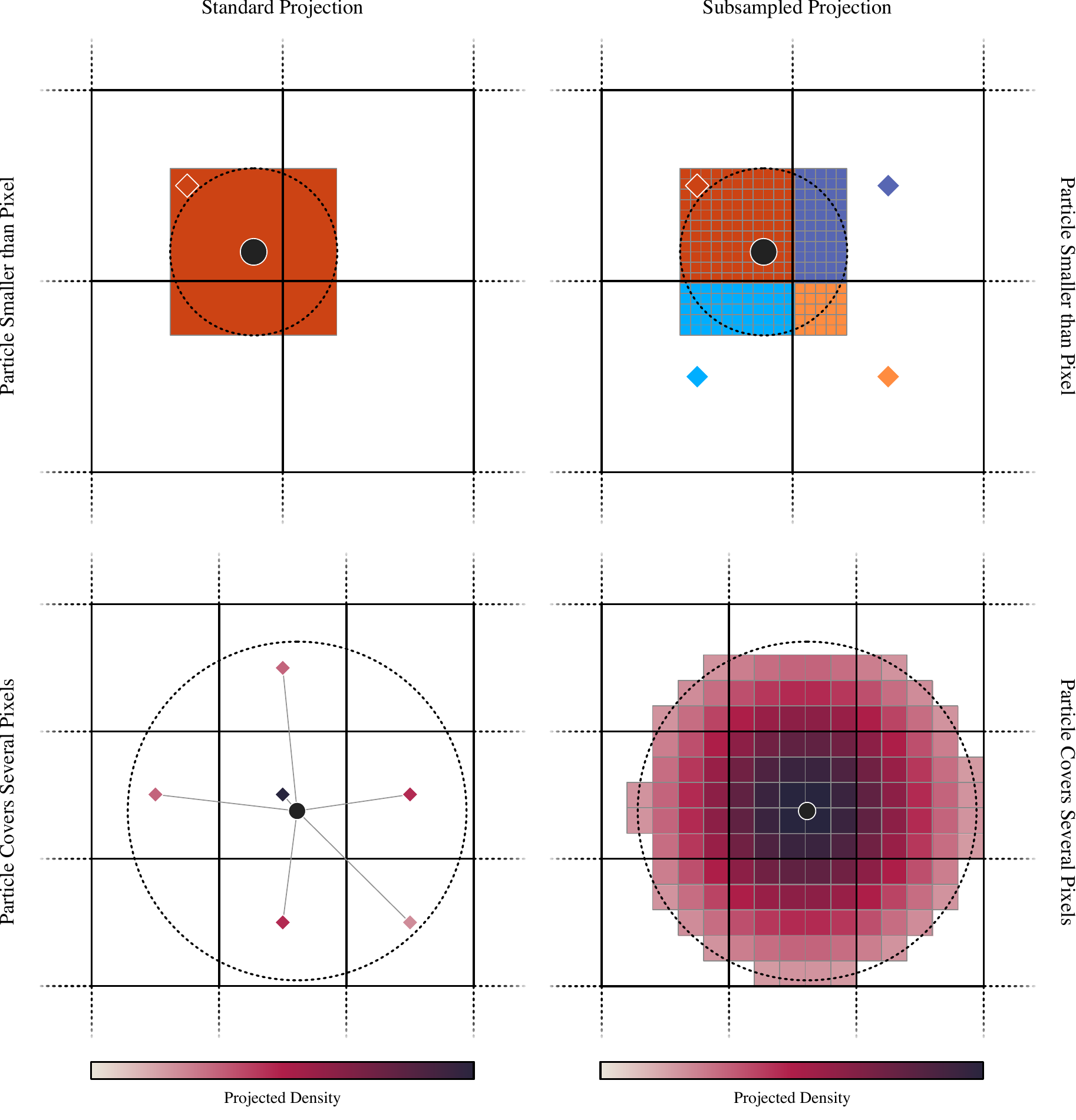}
  \caption{Example cases where the size of the smoothing length (kernel extent shown with dashed lines) of a particle (black point) is smaller than a pixel size (top row) and approximately the size of a pixel (bottom row). The columns show the standard projection algorithm (left) and the sub-sampled, new, algorithm (right). The background black grid shows the pixels, with diamonds representing the kernel evaluation points at the centre of relevant pixels. These are omitted in the bottom right image for clarity, and are replaced with a representation of the sub-pixel grid. In the top left panel, the particle is treated as unresolved, and so its entire density contribution is given to the top left pixel, despite the particle overlapping with three other pixels. In the top right, the blitting grid (grey) is used to appropriately apportion the mass over the four pixels (coloured by their contributing sector of the grid). The bottom left panel shows how the particle usually contributes to each pixel using the evaluated value of the kernel at the pixel centre (with each evaluation point coloured by the appropriate projected density). The bottom right shows the subsampled grid, where the kernel is evaluated at each point to ensure a good reconstruction of the kernel gradient.}
  \label{fig:algorithm}
\end{figure*}

In order to construct converged projections of the field, we must re-integrate our ability to assume that the particles are ordered regularly, and that the field is smooth. To do this, we introduce two new measures.

The first, for particles below the field grid size, is termed `blitting'. Here, we construct a pre-calculated high resolution image of the projected kernel (at $N \times N$ pixels, where here $N=32$). If the kernel extent overlaps any pixel boundary, this pre-calculated kernel is overlaid on the pixel grid and each pixel is assigned the kernel contributions that overlap with the pixel.

The second, termed sub-pixel subsampling (or just `subsampling') ensures that each kernel is well resolved by ensuring that at least $N \times N$ sampling points are made available. If the particle overlaps fewer than $N$ pixels in one dimension, sub-pixels are temporarily generated, which are aligned with the pixel boundaries. This ensures that the kernel gradient is well sampled for all particles. It also ensures that for cases where the kernel does not overlap with the pixel centre, but does overlap with some of the pixel area, that contributions are still made.

A demonstration of these two effects, compared to the standard projection scheme, is shown in Fig. \ref{fig:algorithm}. Algorithmically, the `Subsampled' approach looks like:
\begin{enumerate}
  \item Start with a pixel grid full of zeroes.
  \item Pre-calculate an $N \times N$ grid representing the kernel over a smoothing
        length, and store this.
  \item Take all particle positions $\mathbf{r}_{\rm part}$ and particle
        smoothing lengths $h$. For each particle:
  \item Check if $H< 0.25 {\rm d}x$, and if so:
    \begin{enumerate}
      \item Check for overlaps between this particle and
            neighbouring cells. If there are no overlaps, simply add $m/A$ to
            the containing pixel with $A$ the area of the pixel.
      \item If there are overlaps, use the pre-calculated kernel grid to
            spread the appropriate contribution from this particle over the
            relevant neighbouring cells.
    \end{enumerate}
  \item Otherwise:
    \begin{enumerate}
      \item If this is not the case, figure out how many pixels this kernel
            will overlap $n$. If this is smaller than $N$, sub-sample the pixels
            $N/n$ times (rounded up) at positions $\mathbf{r}_{\rm sub}$.
      \item Loop over the sampling points, and evaluate the projected kernel
            $W(|\mathbf{r}_{\rm part} - \mathbf{r}_{\rm sub}|, h) = W_{\rm sub}$.
      \item Find the mean value of $W_{\rm sub}$ for each pixel, and add this
            contribution to each pixel.
    \end{enumerate}
\end{enumerate}

\begin{figure}
  \centering
  \includegraphics[width=\columnwidth]{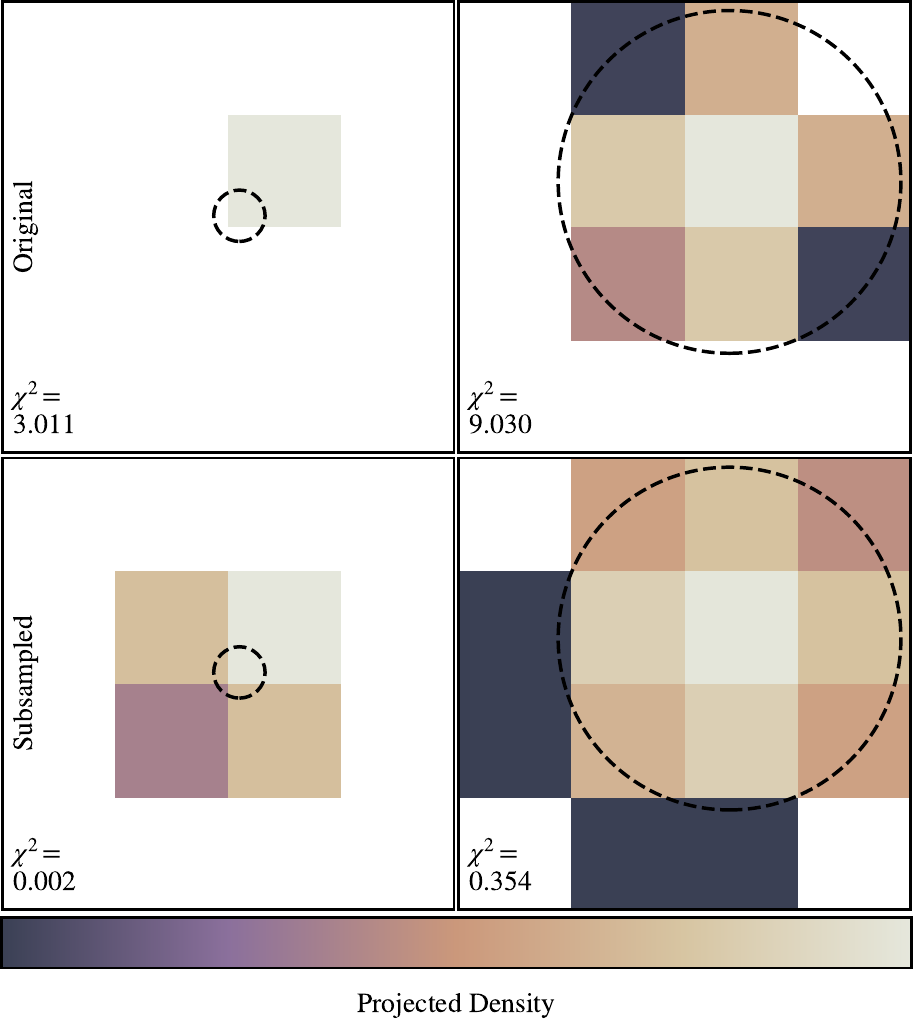}
  \caption{Two pathological cases (columns) corresponding to the pictorial descriptions in Fig. \ref{fig:algorithm}. Again the kernel extent of the particles is demonstrated using a dashed black line. The two rows show how the standard projection technique (`Original') and the `Subsampled' technique handle a case where the particle is much smaller than the $4\times4$ pixel grid used here (left) and only covers a few pixels (right). $\chi^2$ values are given in the bottom left, relative to a very high resolution downsampled grid (1024$^2$). White pixels show where the kernel was never evaluated; in these pixels the projected density is estimated to be exactly zero.}
  \label{fig:pathalogical}
\end{figure}

In Fig. \ref{fig:pathalogical} we demonstrate how the subsampling technique improves performance on two pathological cases on a $4\times4$ pixel grid. The first, shown in the left column, is where a particle with a small smoothing length overlaps a few neighbouring pixels. In the original scheme, the small smoothing length of the particle means all of its contribution is assigned to one pixel, leading to a large error. The `blitting' grid in the subsampled approach ensures that the contributions to neighbouring pixels are correctly captured.

The right column of Fig. \ref{fig:pathalogical} shows a case where the sub-pixel subsampling increases performance by nearly two orders of magnitude. In the original case, the gradient of the kernel in the outer pixels is poorly sampled, leading to an under-estimation of the projected density in these pixels (and in the case of the top-right pixel, no kernel evaluation at all). The sub-pixel subsampling ensures that the kernel is well sampled and that, even in this challenging case, it is possible to reconstruct an accurate projected density field.

If other particles were present in these images, their contributions would lead to a reduction in the overall error. In the top left panel, neighbouring particles would have their contributions assigned entirely to other pixels, perhaps increasing the accuracy of this method. However, this is a case of one error compensating for another, and in an adaptive environment such consistency in the particle arrangement is never guaranteed.
\section{Convergence Properties}

To demonstrate the convergence properties of both the original and subsampled schemes, we use a two dimensional set-up similar to Fig. \ref{fig:gradient}, but aligned diagonally across the pixel grid to further expose issues when assigning the contribution of a whole particle to a single pixel. Here the inter-particle spacing runs from $10^{-3}$ to $10^{-2.5}$ linearly across the space, but now the field is extended so that it covers the second dimension in the plane. The particle positions are then rotated $45^\circ$ so that their spacing is not aligned with the pixel grid.

\begin{figure*}
  \centering
  \includegraphics[width=\textwidth]{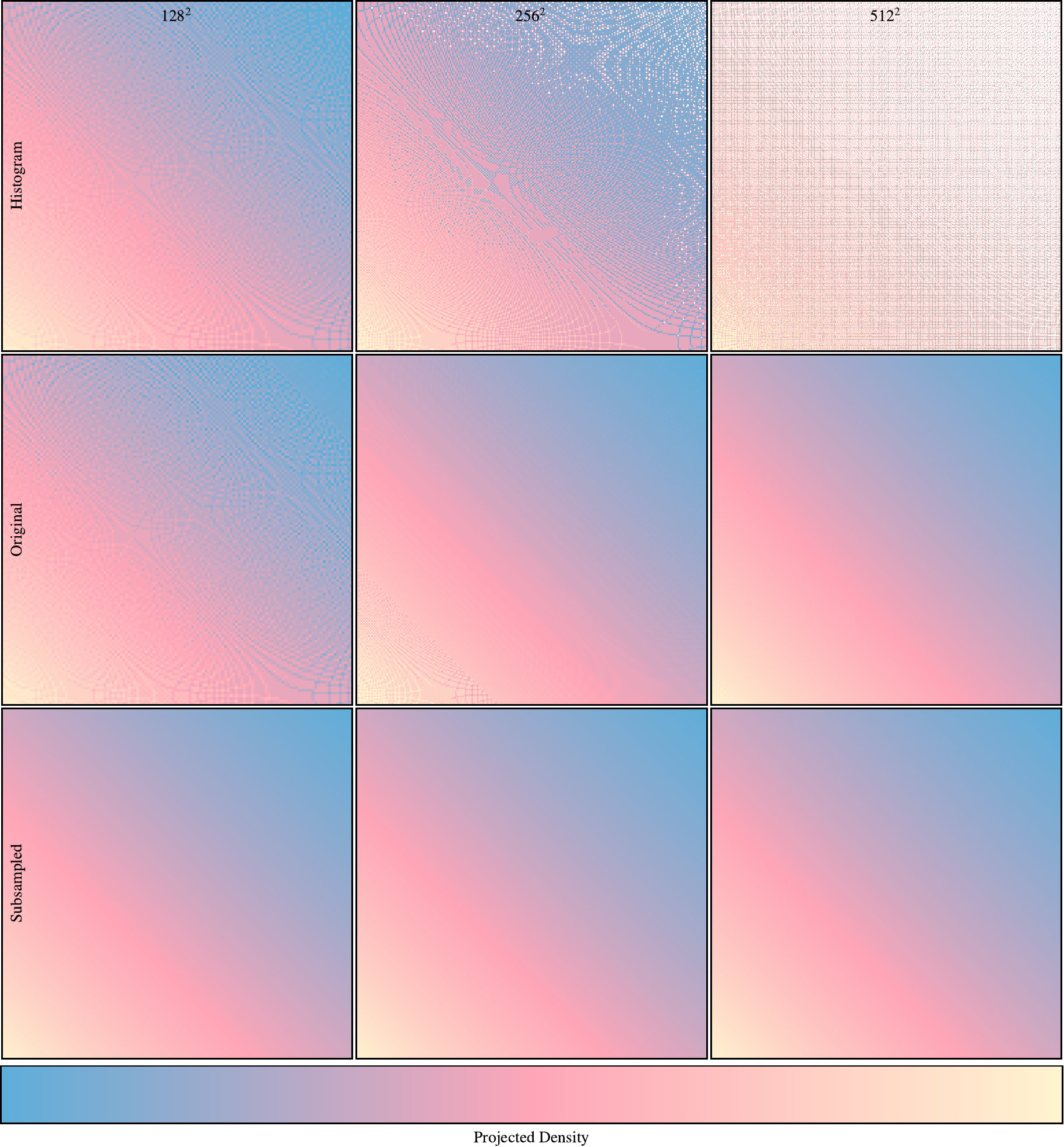}
    \caption{A linear (diagonal) gradient in smoothing length from $10^{-3}$ to $10^{-2.5}$ (in units of the box-size) projected with three different methods (rows) at three different grid resolutions (columns). Here the top method is a simple histogram, where each particle is assigned to its nearest grid centre; this is to demonstrate the basic Moir\'e pattern generated when the grid resolution is close to the particle spacing. The centre row shows how this pattern is also present when using the original scheme for particles below the resolution limit, with the bottom row showing the same gradient for each resolution and demonstrating that the subsampled technique can remove all such patterns. These visualisations are used in Fig. \ref{fig:gradconv} to demonstrate the convergence and costs (Fig. \ref{fig:gradtime}) of each method.}
  \label{fig:gradient2d}
\end{figure*}

The projected densities from this plane are shown in Fig. \ref{fig:gradient2d} using three methods, and three resolutions. The first method, on the top row, only ever uses direct particle assignment to the pixels, and as such is simply a two dimensional histogram. This is provided for comparison with the next two rows that employ the original smoothing scheme and the subsampled scheme. The histogram demonstrates the Moir\'e patterns that occur when the particle distribution and the grid spacing do not align, which would lead to large reconstruction errors.

Those same Moir\'e patterns are shown in the centre-left panel where the original scheme is used with the particle distribution to calculate projected densities on a $128^2$ pixel grid. The subsampled approach, however, does not show any such patterns, mainly due to the use of the blitting grid.

The centre column, calculated on a $256^2$ grid, enables the majority of the particles to have smoothing lengths large enough that the direct particle assignment is not used for the original scheme. However, the poor reconstruction of the kernel gradients in the intermediate smoothing length regime (where $h\approx {\rm d}x$) leads to banding across the opposing diagonal.

\begin{figure}
  \centering
  \includegraphics[width=\columnwidth]{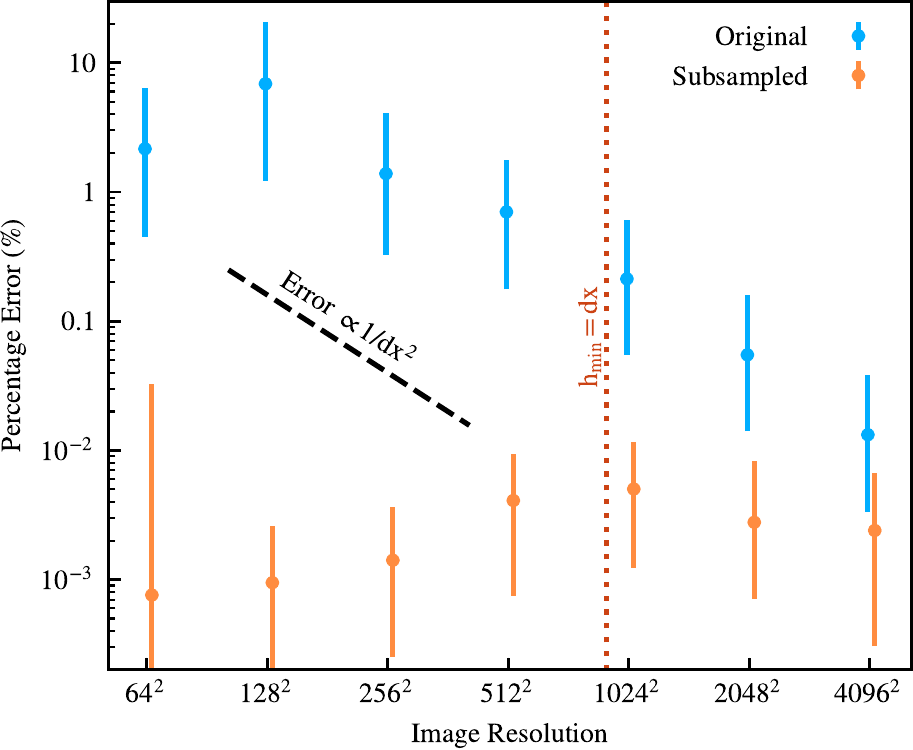}
  \caption{Compares the errors (in the 10\%, 50\%, 90\% percentiles from bottom, point, to top) for the original and subsampled method. Note that each point is evaluated at the number of pixels given on the horizontal axis, with them offset for clarity when the range bars overlap.}
  \label{fig:gradconv}
\end{figure}

In the final column the original approach has visually converged to the subsampled result, with each particle being spread over many pixels. In Fig. \ref{fig:gradconv} we show the numerical convergence, relative to a downsampled $8192^2$ grid.
This is used instead of the analytical solution to remove the impact of any shared systematics between the two approaches, such as the lack of limb brightening in our projected kernel. Percentage error $E_{\%}$ is calculated as
\begin{equation}
  E_{\%} = 100 \cdot \frac{|\rho_{\rm test} - \rho_{\rm ref}|}{\rho_{\rm ref}},
\end{equation}
with $\rho_{\rm test}$ the calculated density reconstruction in that pixel and $\rho_{\rm ref}$ the reference density. Each point shows the median absolute percentage error over all pixels, with the bar showing the 10-90 percentile range.

At all resolutions, the subsampled method ensures that the percentage error in the reconstruction is imperceptibly low, with a highest 90\% value of roughly 0.05\%. Notably, this does not decrease with resolution, as the method is designed to produce converged results at all grid sizes. The original method, on the other hand, demonstrates a high level of error with almost all pixels having at least a 1\% reconstruction error at $128^2$ resolution. This error does indeed converge at the expected rate ($1 / {\rm d}x^2$ as this corresponds to the number of pixels sampling each kernel), but even at the highest resolutions the subsampled approach still provides lower error due to its improved gradient reconstruction of particles with the smallest smoothing lengths (note that $10^{-3} \approx 5 / 4096$, so even at this high resolution the sub-pixel supersampling is still in effect for the highest densities).

\begin{figure}
  \centering
  \includegraphics[width=\columnwidth]{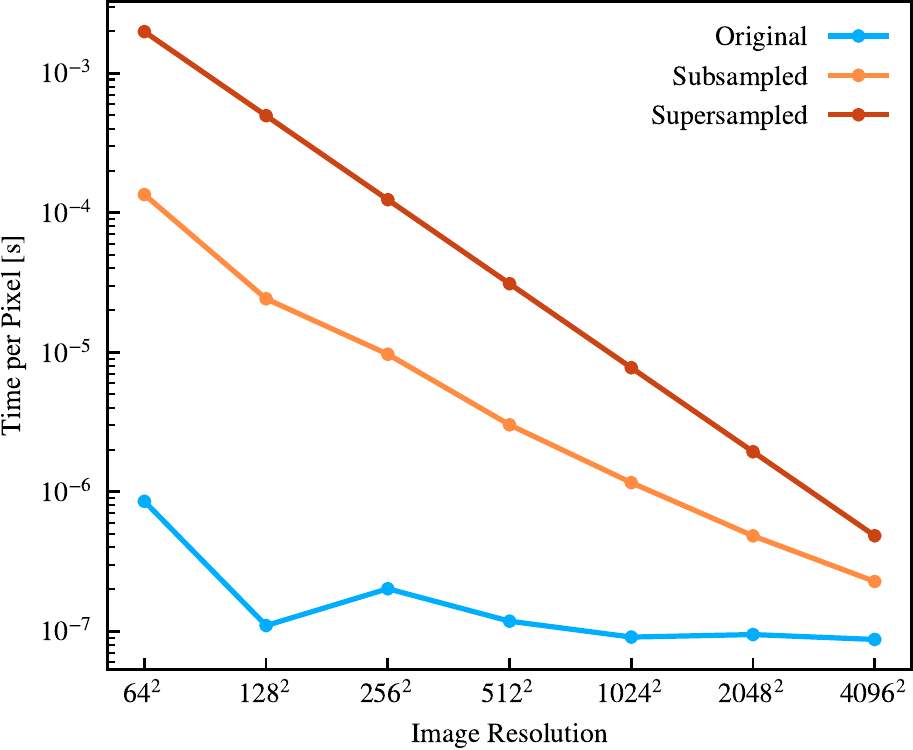}
  \caption{Time-to-solution (using four cores of an Intel i7-8559U) per pixel corresponding to the accuracy results in Fig. \ref{fig:gradconv} for three strategies: the original method with no subsampling (blue); the subsampled method (orange); and using a downsampled pixel grid originally calculated at 8192$^2$ (red). Here we see that although there is significant computational cost associated with the use of the subsampling strategy (along with, of course, the accuracy gains), it still remains approximately 10 times faster than downsampling a large grid.}
  \label{fig:gradtime}
\end{figure}

The improved reconstruction using the subsampling technique hence provides a significant increase in accuracy relative to traditional methods in an adaptive environment. This, however, comes at the cost of increased computing cycles. Fig. \ref{fig:gradtime} shows the scaling of the time to produce the projected density grid for points in Fig. \ref{fig:gradconv}. The red line shows, for comparison, the time taken to compute the $8192^2$ reference grid, which could be downsampled to provide a high accuracy reconstruction. The subsampled technique is still roughly an order of magnitude faster than this, thanks to its efficient handling of particles that are already spread over enough pixels to be well sampled.
\section{Astrophysical Example}

\begin{figure*}
  \centering
  \vspace{-0.5cm}
  \includegraphics[width=\textwidth]{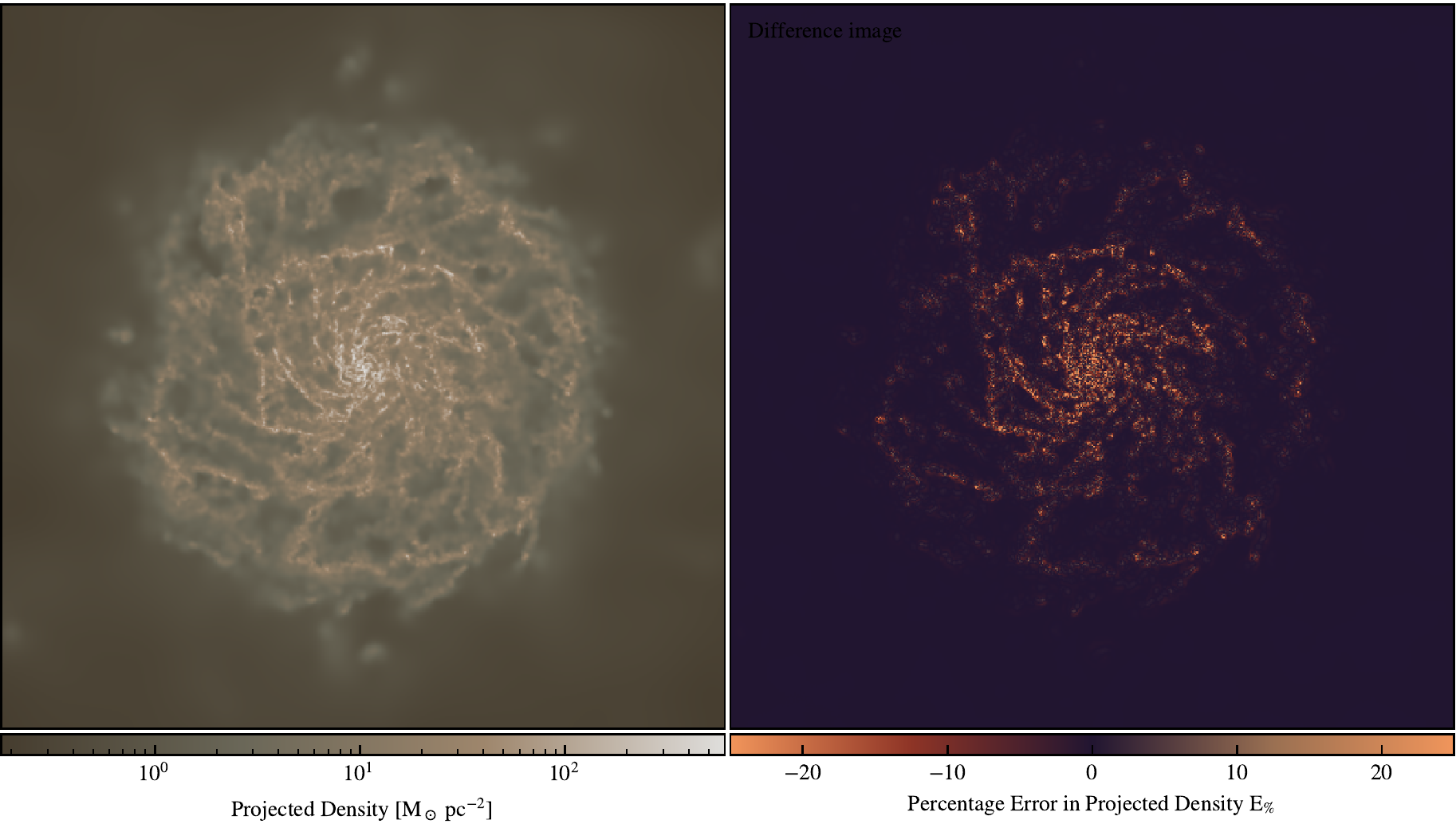}
    \caption{\emph{Left:} Projected (`surface') density of a spiral galaxy from an SPH simulation performed with {\tt SWIFT}. The projected density was calculated on a 512$^2$ grid, with each pixel corresponding to 100 pc of physical space, a common size in observations of real galaxies. \emph{Right:} The error in the estimation of the surface density when not subsampling particles close to, or smaller than, the grid size (i.e. using the `Original' algorithm). This correlates with medium-high density regions of the galaxy, where many particles have smoothing lengths below the grid size. Note that 1 pc = $3.09 \times 10^{16}$ m.}
  \label{fig:error_image}
\end{figure*}

\begin{figure}
  \centering
  \vspace{-0.25cm}
  \includegraphics[width=\columnwidth]{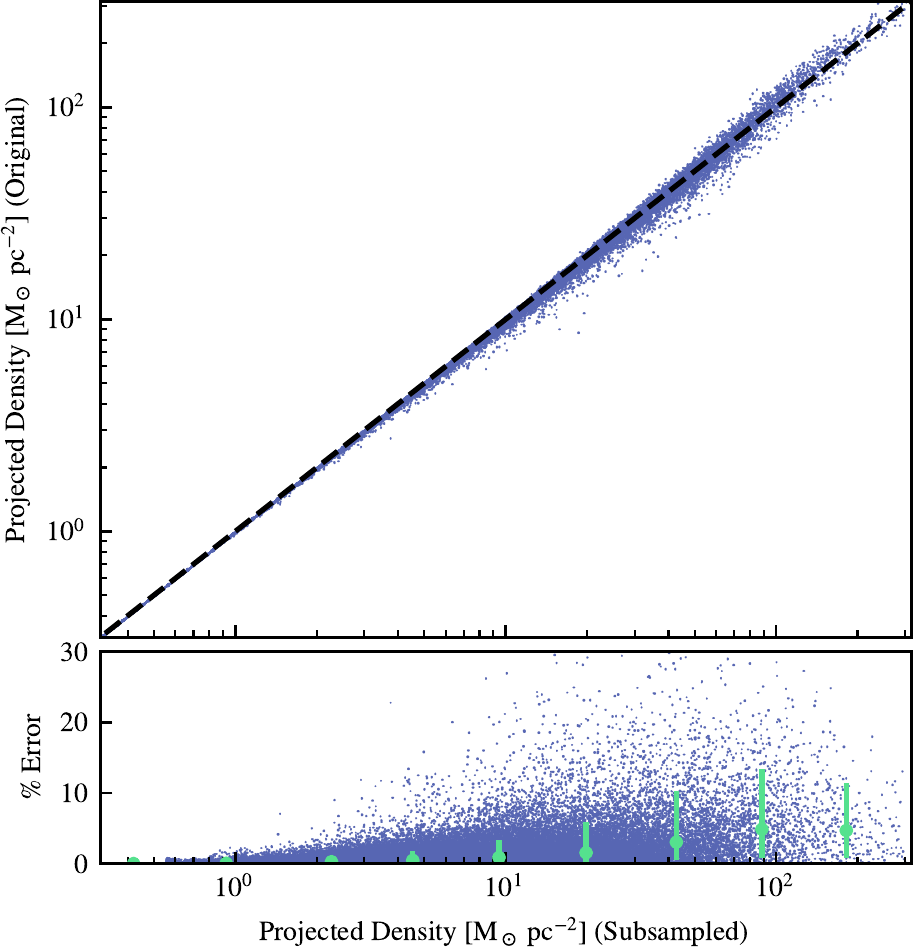}
    \caption{Comparison of projected density calculated with the subsampled strategy (horizontal) and original (vertical) algorithm, for the galaxy in Fig. \ref{fig:error_image}. The black dashed line shows the 1:1 relation. The bottom panel shows the percentage error, with each point corresponding to an individual pixel and the green points corresponding to the 10th, 50th, and 90th percentile (bottom, point, and top of bar) in ten bins across the sampled surface density range.}
  \label{fig:sdcomparison}
  \vspace{-0.15cm}
\end{figure}

So far we have only considered the impact of the subsampling technique on simple problems, but have already demonstrated its ability to provide highly accurate, converged, solutions at all grid sizes. We now consider the astrophysical example of an isolated galaxy disk. The properties of galaxies are the key discriminator between galaxy formation models, as they host an abundance of readily observable properties which can be directly compared with projected quantities from SPH simulations.

This galaxy was simulated with the {\tt SWIFT} code, using the \textsc{Sphenix} \cite{Borrow2020c} hydrodynamics model, and an analytical Hernquist \cite{Hernquist1990} potential for the dark matter component with a mass of $M_{200} = 1.37\times 10^{12}$ M$_\odot$ with M$_\odot = 2\times10^{30}$ kg, concentration $c=9.0$, and a disk fraction of $4\%$. The simulation contains 123533 directly simulated gas particles (of mass $10^5$ M$_\odot$), which interact both through hydrodynamics and gravity, and 424467 star particles (of mass $10^5$ M$_\odot$) that only interact gravitationally. All particles are subject to a modified version of the EAGLE sub-grid model \cite{Schaye2015}, which includes prescriptions for radiative cooling \cite{Ploeckinger2020}, star formation \cite{Schaye2008}, and stellar feedback \cite{DallaVecchia2012}. Here we use a snapshot after $t=0.5$ Gyr, with 1 Gyr $= 3.16\times10^{16}$ s. The initial conditions set-up is described in more detail as the {\tt fid} galaxy in \cite{Schaye2008}. 

Fig. \ref{fig:error_image} shows the projected gas density of the galaxy in the left panel and the percentage error attained from such a projection if using the original scheme. The pixels are set to be 100 pc on a side, for comparison with typical observations of real galaxies (e.g. \cite{Bigiel2008}). The dynamic range in smoothing length here is 300, with $h_{\rm min} = 33$ pc. Here, the percentage error is calculated relative to the subsampled result for more consistent and simple visualisation in later figures, but the result is unchanged when comparing to a high resolution downsampled grid.

The higher density regions of the galaxy (near to the centre) do not have the largest errors due to the stacking of particle layers; this simulation was performed in three dimensions, and as such the Monte Carlo-style sampling is less poor than when only a single layer of particles is considered. In the intermediate density regions, where there is also a strong gradient in density, the error spikes as the kernel gradients are poorly resolved without subsampling. This is a particularly dangerous error structure, as it is typically assumed that the error in SPH simulations decreases as the density increases and the finite mass sampling of the field improves. 

\begin{figure}
  \centering
  \includegraphics[width=\columnwidth]{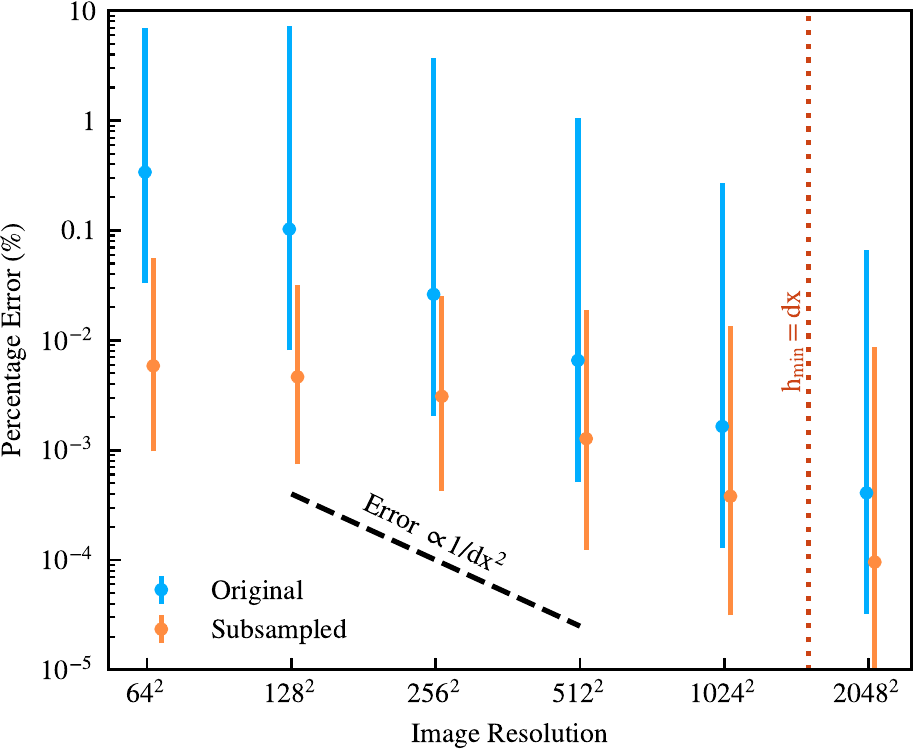}
  \caption{Compares the errors (in the 10\%, 50\%, 90\% percentiles from bottom, point, to top) for the original and subsampled method for the isolated galaxy in Fig. \ref{fig:error_image}. Note that each point is evaluated at the number of pixels given on the horizontal axis, with them offset for clarity when the range bars overlap.}
  \label{fig:galconv}
\end{figure}

The density-dependent error findings are confirmed in Fig. \ref{fig:sdcomparison}, where the projected densities in individual pixels are compared between the two methods. The increase in error with density, except at the highest densities, is confirmed. Additionally, the error is not biased or systematic in any way; the errors inherent in using the poorly sampled projection method are random and as such impossible to offset with, for instance, a scaling factor. 

Fig. \ref{fig:galconv} shows the convergence of the median error with image resolution. Because of the higher dynamic range in this image, relative to the gradient in Fig. \ref{fig:gradconv}, the error ranges for both methods are larger. Here we see that the subsampled method provides a lower than $0.1\%$ error for all pixels, with the original method, as demonstrated in Fig. \ref{fig:sdcomparison}, leading to errors for individual pixels that are higher than 30\%. The median error does converge at the expected rate, but as noted before as real-world observations are taken on coarse grids the convergence properties are not relevant to the accuracy of the method.

These final two figures demonstrate that the correct sampling of kernels in projection is not simply an academic exercise; for simulated observations of real-world problems in adaptive environments subsampling is necessary to produce accurate results.
\section{Conclusions}

In this paper we have demonstrated the issues of using a basic kernel deposition strategy when projecting SPH data to a fixed grid in adaptive scenarios. This strategy was shown to be unable to accurately reproduce a linear gradient in density at low grid resolutions, and also led to large errors when projecting an astrophysical example problem. In addition, these issues are also present (if not exacerbated by even poorer sampling) when constructing the smoothed 3D density at a plane through the field.

To resolve inaccuracies in the deposition of SPH data onto fixed grids, we introduced two techniques:
\begin{enumerate}
  \item Using a pre-calculated kernel grid to `blit' particles that are smaller than the pixel size to ensure that particles that overlap pixels even at this small scale have their contributions weighted accurately.
  \item Using subsampling to ensure that gradients of kernels are always well resolved during the projection process.
\end{enumerate}
With these two techniques we showed that it is possible to produce converged results with very low errors (less than 0.1\%) even when using a low grid resolution. In adaptive scenarios, we hence recommend the use of subsampling techniques.

\section*{Acknowledgment}
The authors thank Matthieu Schaller and Alejandro Benitez-Llambay for helpful discussions.
They also thank Loic Hausammann for contributing a GPU-parallel version of the visualisation algorithms (not used in the prior examples) to the library.
JB is supported by STFC studentship ST/R504725/1.
AJK acknowledges an STFC studentship ST/S505365/1.

\IEEEtriggeratref{19}


\vspace{3.5mm}
\bibliographystyle{IEEEtran.bst}
\bibliography{IEEEabrv,bibliography}
%

\end{document}